# Coaxial Jets and Sheaths in Wide-Angle-Tail Radio Galaxies

D. M. Katz-Stone

United States Naval Academy, Annapolis, MD 21402-5026

L. Rudnick

University of Minnesota, Minneapolis, MN 55455

C. Butenhoff[1]

University of Minnesota, Minneapolis, MN 55455

and

A. A. O'Donoghue

St. Lawrence University, Canton, NY 13617

[1]present address: University of Calgary, Deptartment of Physics and Astronomy, 2500 University Dr. N. W., Calgary, Alberta, T2N1N4 Canada

# ABSTRACT

We add $\lambda\lambda$ 20, 6 and 3.6 cm VLA observations of two WATs, 1231+674 and 1433+553, to existing VLA data at $\lambda\lambda$ 6 and 20 cm, in order to study the variations of spectral index as a function of position. We apply the spectral tomography process that we introduced in our analysis of 3C67, 3C190 and 3C449 (Katz-Stone 1995, Katz-Stone and Rudnick 1997a, Katz-Stone and Rudnick 1997b). Both spectral tomography and polarization maps indicate that there are two distinct extended components in each source. As in the case of 3C449, we find that each source has a flat spectrum jet surrounded by a steeper spectrum sheath. The steep components tend to be more highly polarized than the flat components. We discuss a number of possibilities for the dynamics of the jet/sheath systems, and the evolution of their relativistic electron populations. While the exact nature of these two coaxial components is still uncertain, their existence requires new models of jets in FR I sources and may also have implications for the dichotomy between FR Is and FR IIs.

*Subject headings:*

## 1. Introduction

This paper centers upon two radio galaxies that are classified as Fanaroff-Riley type I (FR I, Fanaroff and Riley 1974) because they are limb darkened, and as Wide Angle Tails (WATs) because of their shallow C shapes (Owen and Rudnick 1976). Our original interest was in the variation of spectral index in these sources as a function of distance from their cores. Spectral variations are a key diagnostic for extragalactic radio sources, being used to derive such quantities as particle ages and flow speeds, and to identify possible sites of relativistic particle acceleration. Since WATs are normally considered to consist solely of *outflowing* material, bent by some external interaction (e.g., Eilek et al., 1984, Loken et al., 1995, Burns et al., 1994a,b ), spectral variations produced by the relativistic particle aging should be separable from other effects.

In order to measure spectral variations, we add here Very Large Array (VLA[2]) observations of the WATs 1231+674 and 1433+553 to the existing data of O'Donoghue et al. (1990). Our original intent was to add a third frequency to measure the shapes of the spectra as a function of position in the sources (Katz-Stone et al., 1993). These sources are particularly interesting because previous spectral index data showed an initial steepening of the spectrum with distance from the nucleus followed by a flattening of the spectrum in the brighter, or hot spot regions. Our goal was to look for possible changes in the spectral shape –an indication of *in situ* particle acceleration – as the jet passed through the hot spot.

As explained below, we were unable to use the existing $\lambda$ 6 cm data for spectral shape measurements. However, we find that our original goals could not have been fulfilled even

---

if we had multifrequency data. Our analysis shows that there are two partially overlapping spectral components in each source, and neither of these components individually show the reported steepening of the spectrum followed by a flattening. We then focussed our attention on the critical issue of the nature and relationships of these two components and their implications for such questions as the history of the relativistic particles and the dichotomy between FR Is and FR IIs.

In this paper, we will use the term "sheath" to describe a lower surface brightness feature surrounding a jet. A number of other terms are found in the literature, but they usually carry model-dependent or multiple meanings. Bridle and Perley (1984), in their review of extragalactic jets, cite the existence of sheaths in a variety of sources, and note that it is unclear whether these are "outer jets, static sheaths, or backflows." That dilemma remains today.

The first extensive discussion of a sheath is that by O'Dea and Owen (1987) for the head-tail source NGC 1265. They detect a pair of sheaths only in their longest wavelength ($\lambda$ 21cm) map and infer that they must have steeper spectra than the jets. O'Dea and Owen (1987) discuss various possible origins for the sheaths, which are tens of kpc long, and conclude that in NGC 1265 the diffuse emission is not part of the outflowing jet, but a result of particles that have diffused out of the jet into a static external medium.

Laing (1993, 1996) invokes a jet/sheath structure, with a relativistic jet, to explain a wide variety of statistical results including the brightness asymmetries, polarization structures and depolarization, gap sizes, and core prominences, for jets up to scales of several kpc. Although the slower moving sheaths can usually not be isolated on these scales, their presence is convincingly demonstrated in the statistical trends.

On the physical side, there are two major classes of jet/sheath models. One is the formation of a distinct boundary layer on a jet, e.g., through turbulent entrainment of external material (e.g., Bicknell 1996). The other is a two-component jet flow, produced in the nuclear regions, such as an electron-positron jet from a black hole, with a slower electron-proton wind from its surrounding accretion disk proposed by Hanasz and Sol 1996. In Section 4, we will briefly comment on the relationship between our observations and such models, although a great deal of work remains to be done.

A further consequence of the jet/sheath structure in the WATs found here and in 3C449 is how they relate to the classical division of extragalactic sources into the structurally defined FR I and FR II categories. FR II sources, with hot spots at their extremities, are often assumed to be powered by supersonic jets which then form post-hot spot backflows, while FR I sources, without external hot spots, are thought of as outflowing jets which become subsonic, e.g., through turbulent entrainment and broadening (e.g., De Young 1984, Norman et al., 1988). It is not clear whether the WAT sheaths are inflowing or outflowing material. In addition, the WATs have monochromatic radio luminosities near the transition between FR Is and FR IIs ($2 \times 10^{25}$ W Hz$^{-1}$ sr$^{-1}$ at 178 MHz; Fanaroff and Riley 1974.) The corresponding properties for 1231+674 and 1433+553 are summarized in Table 1, along with data for 3C449, whose analysis was published previously (Katz-Stone and Rudnick 1997b.) In Section 4, we will therefore discuss the possibility that WATs – with their sheaths of unknown origin – may represent transition objects between the outflowing FR Is and the backflowing FR IIs.

EDITOR: PLACE TABLE 1 HERE.

## 2. Observations

We observed 1231+674 and 1433+553 with the VLA at 3 frequencies: $\nu = 1445.0$, 4525.0 and 8435.11 MHz ($\lambda\lambda = 20$, 6 and 3.6 cm) as summarized in Table 2. We added these to previous data from O'Donoghue et al. (1990) at $\lambda\lambda$ 20, 6 cm.

EDITOR: PLACE TABLE 2 HERE.

The primary flux and polarization angle calibration for all wavelengths was based upon 3C286. The assumed flux densities were set to the Baars' flux scale (Baars et al., 1977), using the values incorporated into the AIPS task SETJY by R. Perley, and listed in Table 3 here.

EDITOR: PLACE TABLE 3 HERE.

All the data were calibrated using standard techniques with 1311+678 (1418+546) as phase calibrator for 1231+674 (1433+553). In addition there were 20 to 30 self-calibration passes of phase calibration for each source. All the maps in this paper were deconvolved with the 'CLEAN' algorithm as implemented by the AIPS task 'MX' and were corrected for the attenuation due to the primary beam. For the comparisons between maps of different frequencies, all maps were convolved to a common resolution of $\approx 2.''8$.

Since these data were self-calibrated, absolute positional information is lost. Therefore, there could be a small relative position shift between the various maps. Based upon our examination of spectral features in 1433+553, we determined that a shift of the $\lambda$ 3.6 cm map (0.''15, 0.''16) in (RA, Dec) was necessary to properly align it with the $\lambda$ 20 cm map. No shift was necessary or performed on 1231+674.

The $\lambda$ 6 cm maps of both sources showed very faint ($\approx 1.5$ times the off-source rms noise) ghost-like images of the sources superimposed slightly to the east of the normal image. These only became apparent when we performed our detailed spectral analysis. This ghost feature created an apparent gradient in the spectral index between $\lambda\lambda$ 6 cm and 3.6 cm along the east-west axis in 1231+674 such that the spectral index appeared to drop from $\approx$ +0.5 to -0.5 from east to west across both tails. These unusual features did not appear in the $\lambda\lambda$ 20 or 3.6 cm maps or in low resolution $\lambda$ 6 cm maps using the new data alone. We made a number of attempts to understand the cause of these ghost-like features. We examined each IF, polarization, and observing date separately to see if the ghost features were isolated. Because the O'Donoghue et al. data were taken twelve years ago, we tried self-calibrations with both the modern AIPS task CALIB and the older task ASCAL. We also tried to vary the parameters of self-calibration; we used both a point source and the $\lambda$ 20 cm data as models. In addition, we used the maximum entropy deconvolution method employed by the AIPS task VTESS to ensure that the ghost images were not an artifact of MX. Because the data were taken at separate times and with slightly different frequencies, DBCON was used to merge the datasets and scale the uv coordinates to a common frequency. We verified that no errors were introduced in this step by calculating several points by hand. None of these efforts was successful at identifying the origins or eliminating the ghost images, so we have not included the $\lambda$ 6 cm data in the spectral analysis. However, the $\lambda$ 6 cm polarization data are still useful, since errors of a few percent in the polarized signal would not be significant.

## 3. Results

### 3.1. Total Intensity

The total intensity maps at $\approx 1''$ resolution are shown in Figures 1 and 2 and the $\approx 2.8''$ maps in Figures 3 and 4. We can see that each source has a pair of jets originating near the nucleus, and a region of rapid widening at distances of $\approx 20\ kpc$ from the cores. Such behavior is commonly ascribed to a broadening jet, and our lower resolution maps could be considered consistent with that picture. However, the higher resolution maps and the tomography gallery (as discussed below) make it clear that there are two components present, a jet and a sheath. It is the appearance of the sheath that is responsible for the apparent broadening. In both of these sources the sheaths themselves appear to expand, or flatten before they terminate.

EDITOR: PLACE FIGURE 1 HERE.

EDITOR: PLACE FIGURE 2 HERE.

EDITOR: PLACE FIGURE 3 HERE.

EDITOR: PLACE FIGURE 4 HERE.

## 3.2. Spectra

We introduced the spectral tomography gallery as a tool for separating spectral components (Katz-Stone 1995, Katz-Stone and Rudnick 1997a, Katz-Stone and Rudnick 1997b). A gallery consists of a series of tomography maps where $\alpha_t$ is varied. Each map is calculated as

$$I_{tom}(\alpha_t) \equiv I_{20} - \left(\frac{\nu_{20}}{\nu_{3.6}}\right)^{\alpha_t} I_{3.6} \qquad (1)$$

A feature of spectral index $\alpha_{20}^{3.6}$ disappears from the map in which $\alpha_{20}^{3.6} = \alpha_t$. Figure 5 and 6 are the tomography galleries for 1231+674 and 1433+553 respectively. Both sources show two distinct spectral components.

The bright inner 1.'2 of 1231+674 shows a flat component embedded in a steeper component in the tomography maps for $\alpha_t = -0.55$ to $-0.75$, reminiscent of 3C449 (Katz-Stone 1995, Katz-Stone and Rudnick 1997b). We will refer to the northern (southern) half of this bright 1.'2 region as N (S). Further, we will call the flat component in both regions the "flat jet" and the steep component the "sheath." In addition, the more diffuse regions farther out show a similar two component structure in the $\alpha_t = -1.2$ tomography map.

Similarly, 1433+553 appears to have two overlapping spectral components. This is best seen in the $\alpha_t = -0.65$ to $-0.75$ tomography maps. In the north, the two components are identifiable only until the bend, and are somewhat difficult to see in this display. We will refer to the region before the bend as N and the region after the bend as NW. The last two tomography images ($\alpha_t = -0.95$ and $-1.15$) show a spectral steepening with distance from the bend in NW.

EDITOR: PLACE FIGURE 5 HERE.

EDITOR: PLACE FIGURE 6 HERE.

We measured the spectral indices of the two separate components as a function of distance from the core. To estimate the spectral index of the flat jet at a particular position, we made slices at that position through a series of spectral tomography maps. In Figure 7, we show an example of this procedure for 1433+553. The flat jet is at a position of +2 arcsec along this slice. By searching the spectral tomography slices at each location for the one in which the flat jet vanishes with respect to the sheath, we determine the spectral index of the flat jet at that position. The spectral indices at various points along the sheaths were determined using slices in the spectral index maps. We used points where the sheath was as far as possible from the flat jet, and therefore least contaminated, but also still had sufficient signal:noise. For some positions, the separation was on the order of the resolution; therefore the actual spectral index of the sheath may be steeper than our estimates (Figures 8 and 9). Where the sheath was separable and different on both sides of the flat jet, we found the spectral index of the sheath for both places. As a final note about this process, we have assumed that where the components overlap, each has a simple profile; if there is substructure in one or both of these components our estimate of the spectral index is a combination of the spectral index of the substructure and the actual component.

EDITOR: PLACE FIGURE 7 HERE.

Figure 8 and 9 are plots of the spectral indices of the two components (flat jet and sheath) as a function of distance from the core for 1231+674 and 1433+553, respectively. The uncertainty in the spectral index of the flat jet comes from the uncertainty in determining in which spectral tomography slice the flat jet vanishes; the uncertainty in this procedure is greater than the error from the off-source rms noise.

These plots reveal several important facts about the spectral indices. The spectral indices for the flat jet and the sheath are clearly different for both sources. Within 1231+674 N and S (about 65 kpc from the core) there is little variation in the spectral index of either the flat jet or the sheath. Beyond 1231+674 N and S, both the flat jet and the sheath steepen. In 1433+553 S the spectral indices of the flat jet and the sheath are constant, and only small variations are seen in the N and NW components. In the sheath there is a discontinuity in the spectral index in going from N to NW (about 55 kpc from the core) which may be due to our inability to cleanly separate the sheath from the flat jet in the N, making the sheath appear flatter than it truely is. Similarly, in NW the flat jet is steeper than it is in N; this may be due to our inability to cleanly separate the flat jet from the sheath. In Figure 8 and 9, most of the spectral indices derived by O'Donoghue et al. (1990) fall between the flat jet and the sheath. Their measurements show a change in spectral index as a function of position. This is really a reflection of which component (the flat jet or the sheath) dominates at that position; for example, the dramatic steepening of the spectrum observed by O'Donoghue et al. (1990) between 40 and 50 kpc in 1433+553 S does not correlate to a *change* in the flat jet or the sheath (Figure 9).

EDITOR: PLACE FIGURE 8 HERE.

EDITOR: PLACE FIGURE 9 HERE.

### 3.3.  Polarization

Polarization maps for 1231+674 and 1433+553 are shown in Figures 10 and 11 respectively. The amplitudes of the vectors show the polarized flux at $\lambda$ 3.6$cm$, while their directions show the calculated direction of the *magnetic* fields. The noise in the polarized intensity maps was 30$\mu$Jy. The magnetic field directions are defined to be 90° from the Faraday de-rotated electric fields. The Faraday rotations were measured from $\lambda$ 3.6 $cm$ and $\lambda$ 6 $cm$, where the difference in polarization angle between the two frequencies was forced to be less than 90°. This yielded a range of rotation measures with mean (rms) of -58 (290) $\frac{rad}{m^2}$ for 1231+674 and -19 (240) $\frac{rad}{m^2}$ for 1433+553. Characteristic errors on the magnetic field direction plots are ±15°. Adding the $\lambda$ 20 $cm$ data was not useful because the $n\pi$ ambiguities at different locations could not be resolved.

EDITOR: PLACE FIGURE 10 HERE.

EDITOR: PLACE FIGURE 11 HERE.

The fractional polarizations are highest on the edges of the sources, reaching values above 50%. The magnetic fields have considerable structure. In 1231+674 S, the fields are along the direction of the inferred flow, except for the faint southernmost extremity, where they become transverse. In 1231+674 N, the same is true, with the addition of a transverse region on the flat jet, at the position where the sheath first appears. In 1433+553 N, the field is also along the apparent flow direction. In 1433+553 S, the vectors tend to be along the flow line, except at the southwest end, where the field keeps the same direction, but the source appears to turn by 90°.

Figure 12 is an overlay of the polarization intensity at $\lambda$ 3.6 cm with a spectral tomography map for 1231+674, and Figure 13 is an overlay of the fractional polarization at $\lambda$ 3.6 cm with a spectral tomography map for 1433+553. The polarization intensity is generally higher near the edges of 1231+674 and lower in the center. In 1433+553, the fractional polarization is high along NW and the outer edges of S. There is not a one-to-one correspondence between the spectral components and the polarization, but the flat jet tends to be less polarized than the steep spectrum sheath.

EDITOR: PLACE FIGURE 12 HERE.

EDITOR: PLACE FIGURE 13 HERE.

## 4.    Discussion

The existence of sheath-enclosed jets reported here and in other sources raises a number of important questions regarding their formation, dynamics, and the evolution of their relativistic plasmas. In this section we explore some of these issues, and their possible relationship to the dichotomy between FR I and FR II sources. We will not explicitly resolve the key questions that we raise, but we will discuss what sort of research is needed.

The first class of questions has to do with origins: Is the sheath formed along with the jet in the nuclear regions? Or does it develop much later on scales of kpc? Coupled to these questions are those of dynamics: Is the sheath outflowing? Or static with respect to the ISM? Or backflowing, as happens in FR II, high luminosity, sources? What speeds and densities are involved? Other issues involve the relativistic plasma: What is the origin of the magnetic fields and relativistic electrons? Are there two independent populations or one? How do these evolve with time? Is there ongoing particle acceleration as well as losses? How are these coupled to the source dynamics?

## 4.1.  The Structure and Dynamics of Jets/Sheaths

On the question of origins, we have no information. This would require observations on pc scales at more than one frequency, with sufficient information to resolve the jet(s) transversely. This is a challenging but worthwhile task, to identify whether the sheaths are present even on scales less than a kpc, or whether their sudden appearance and broadening on scales of $\approx 10$ kpc in the WATs indicates their actual onset. The recent observations of Baum et al., 1997 suggest the possible presence of a sheath in 3C264 at scales of a few hundred pc, and further such detailed studies are needed.

Our observations do begin to offer insights into dynamical questions. We find that the jets continue, with little or no observed broadening, after the appearance of the sheaths. It is not clear whether this behavior is consistent with the essentially one-dimensional models of turbulent-broadened jets (*e.g.*, Bicknell 1984, Komissarov 1994). Narrow jets in backflows and intrinsically two-component outward flows (*e.g.*, Sol et al., 1989, Hanasz and Sol 1996) can be relatively stable configurations, and thus are attractive models for the jet/sheath phenomenon. Narrow jets are also found to persist along with the formation of a more diffuse structure after crossing shocks (*e.g.*, Loken et al., 1995), so this may be a promising alternative to explore further.

Flow velocities are a key physical parameter of radio galaxies. Because jets are assume to represent outward flow, in theory, their flow velocities can be estimated by determining the age of the relativistic particles and the distance over which they have traveled. The distance traveled is, of course, derived from knowing the distance to the source, and the age of the particles comes from $t \propto \frac{1}{\sqrt{(B^3 \nu_{brk})}}$. The break frequency, $\nu_{brk}$ is the frequency at which synchrotron losses becomes significant. This is derived by 1) measuring the shape of the synchrotron spectrum (including the injection spectral index) which requires at least three observed frequencies (Katz-Stone et al., 1993) and 2) confirming that this shape does not change which would indicate *in situ* particle acceleration. The magnetic field strength, B over which the particles have traveled could be estimated from the emissivity and depends on such assumptions as an equipartion of energy.

On the other hand, given the uncertain nature of the sheaths (outflow? backflow? static?) it is currently not possible to calculate their velocities on dynamical grounds. However, the polarization maps along with numerical models provide some insight. The magnetic fields in the sheaths observed here are aligned with the apparent flow directions. This is expected both in models with turbulent, sheared boundary layers, (*e.g.,* Killeen and Bicknell 1988) as well as in two-component flows where axial fields provide a more stable configuration (Hanasz and Sol 1996). It is also consistent with the sheaths being backflows, by analogy with the flow-aligned fields in FR II lobes. At the ends of at least 1231+674's tails, the fields become transverse to the direction from the core. This would be expected if the flows were actually brought to a halt at this point, compressing the magnetic fields. This is very different behavior than expected from more extreme FR Is, where the outward flow is not brought to a halt, but simply decelerates gradually. Finally, the higher fractional polarization apparent in the steep spectrum sheaths may also be indicative of the flow process. However, the effects of the various flow processes on the percent polarization have not yet been worked out in detail.

In concluding this section, we note again that at present, we do not know if the sheaths in these WATs represent material that is outflowing, backflowing, or static with respect to the jets. Resolving that dilemma will involve much more detailed study of sheath dynamics and relativistic plasma evolution, separate from the contaminating influence of the jets.

## 4.2.   Independent Relativistic Particle Population Models

One of our key findings relates to the distribution of relativistic particles. We found that although there is a significant difference in the spectral indices *between* the flat jet and the sheath, there is comparatively little variation in index *within* each spectral component. This is most dramatic over the first 30 kpc (for 1231+674) or 60 kpc (for 1433+553) where the jet and sheath are seen to overlap.

One possibility to explain these results is that the relativistic electron populations in the jets and sheaths are truly independent. This could arise in a variety of ways. The jet and sheath themselves could be independent flows, and whatever processes originally populated them with relativistic particles did so differently in the two components. Another possibility is that there is ongoing *in situ* particle acceleration, either in the jet or sheath, or both, under different physical conditions. Alternatively, a second electron population could originate in the ISM/ICM, and become lit by the disturbances created by the passing jet (T. W. Jones private communication 1995). In order to resolve the possibilities, we need observational information on the low-frequency power-law indicies separately for the jets and sheaths, and theoretical models that include the evolution of relativistic particles.

Another possibility is that the sheath represents an older epoch of activity, while the jet illuminates the currently energized flow. A variety of intermittent source models have been proposed for other reasons (*e.g.*, Burns, Christiansen and Hough 1982, Rudnick and Edgar 1984, Reynolds and Begelman 1997). Another interesting model is described in the review article by Begelman et al. (1984), who describe a scenario in which the jet of an FR II is suddenly turned off. The evacuated jet channel would then be filled in transversely by sheath material. The collapsing sheath would be pear-shaped with the smaller end point towards the nucleus. If the sheath material falls in at supersonic speed, shocks would form and there would be particle acceleration along the evacuated jet. This could then account for the relatively constant, flat spectral index that we observe in the flat jets of 3C449, 1231+674 and 1433+553 as well as the general pear-shape morphology. In such a model, the jet does not represent outflowing material, and velocity estimates are irrelevant.

In the context of multiple-epoch models, we note a curious resemblance between the high and low brightness level structures in the southern half of 1231+674. In Figure 1, just south of Dec 0", we see the sheath-engulfed jet pointing due south and then turning/widening mostly westward at about Dec -20". The sheath itself appears to have a similar structure, but on a larger scale. From just south of Dec 0" to about Dec -40", the sheath mainly points southward; just south of Dec -40" it turns/widens towards the west. While the possibility of separate electron populations, is tantalizing and cannot be ruled out by our current observations, in the next section we will explore the possibility of a single electron population.

### 4.3. Single Relativistic Particle Population Models

Another class of possibilities is that the sheath and flat jet started out with the *same* population of relativistic electrons, and that these electrons were subsequently separated by some dynamical process, such as diffusion out of the jet into the surrounding medium, or the splitting off of some jet material by shear and turbulence. The zeroth order demonstration of this would be to see that the sheath has the same shape spectrum as the flat jet; that one spectrum had merely been shifted along the log $(\nu)$ and log(I) axes with respect to the other. After establishing that the sheath and the flat jet have the same spectrum, we would look for physical conditions that could account for the shifting of one spectrum with respect to the other spectrum. If the spectral shapes were different, then we would conclude there *are* two separate relativistic electron populations, and we have outlined the consequences of that above.

If the *shape* of the electron energy distribution is the same in the flat jet and the sheath, then the steeper spectral index in the sheath must be due to a lower value of the break frequency ($\nu_{brk} \propto \gamma_{brk}^2 B$) where electrons with energies greater than $\gamma_{brk}$ have suffered significant radiative losses. A lower value of $\nu_{brk}$ in the sheath can then occur in one of two ways – either the magnetic field is lower or the relativistic electrons in the sheath may have lost energy through expansion or radiation. The magnetic field in the sheath may simply be lower than in the flat jet or the sheath may be an adiabatically expanded version of the flat jet. In these cases the spectra would be shifted along the log($\nu$)and log($\nu$) axes by equal amounts such that

$$\frac{\nu_{sheath}}{\nu_{jet}} = \frac{I_{sheath}}{I_{jet}} = \frac{B_{sheath}}{B_{jet}}$$

where $\nu_{component}$ is the break frequency of that component and comes from determining the shape of the spectrum as stated above. Therefore, the first order test of these scenarios is to see if the sheath spectrum can be overlayed with the jet spectrum by sliding it by equal amounts (as determined from the ratio of the break frequencies) along the $\log(\nu)$ and $\log I$ axes. To further confirm and test these results would involve estimating the magnetic field in each component independently from the change in break frequency, e.g. by assuming local equipartition of energy.

Another possibility is that the sheath is older than the flat jet. As described above, age calculations require 1) measuring the shape of the spectra, 2) determining $\nu_{brk}$ and 3) estimating the magnetic field. A further consistency check should be performed to ensure that the emissivity has fallen off in accordance with a standard aging model (e.g., Leahy 1991.)

To test the plausibility of these various models, we first made a series of simplifying assumptions about the relativistic electron population in the flat jet and sheath. We assumed they had the Jaffe and Perola (Jaffe and Perola 1974) spectral shape which allows for pitch angle scattering, a low-frequency (injection) spectral index of -0.5, and an equipartition of energy with the magnetic field. We then estimated the ratios between the flat jet and sheath in surface brightness, size and emissivity, break frequency, and calculated magnetic field strength. We find that there is no self-consistent explanation for the observations using a single jet/sheath electron population, although this is not a strong conclusion because of large uncertainties in the estimated break frequencies.

Alternative models with electrons that have experienced different radiative losses (aging) in the jet and sheath also appear inconsistent with the data. The lower inferred magnetic fields in the sheath should have resulted in flatter spectra, the opposite of what is observed. The spectra of the flat jet and sheath should also show monotonic steepening with distance from the nucleus, but neither is observed to change for 10s of kpc. If the electrons are thought to diffuse from the flat jet and lose energy, then a transition region also should have been observed between the young and old populations. Instead, Figures 8 and 9 do not show a smooth transition from young electrons to old electrons; rather there is dramatic jump in apparent age between the two populations. This raises several questions. Where and how did the older population age? Do two different populations emerge from the nuclear region? Why is there little aging within either the flat jet or the sheath alone?

We note that more than one of the above processes may be important; realistic models of jet/sheath flows with relativistic particles will be necessary to resolve these questions.

## 4.4.    Sheaths and the FR I/II Dichotomy

As discussed above, the most fundamental flow properties of these WAT sheaths are uncertain. If one were to ignore the emission beyond the bright regions (hot spots), the structures would appear similar to FR IIs, with the sheaths corresponding to the steeper spectrum backflow cocoons. But the emission beyond the hot spots makes them similar to FR Is, which in the extreme case of head-tail sources such as NGC1265, are clearly completely outflowing material. Nor do our spectral results provide guidance in this area, since the gradients between the cores and hot spots are small and could easily occur through magnetic field variations. We thus find that these WATs contain properties of both FR I and FR II sources, and perhaps should be considered intermediate between them.

The intermediate nature of these WATs is also consistent with their luminosities. As shown in Table 1, 1231+674, 1433+553, and 3C449 are all within an order of magnitude of the $\approx 2$ x $10^{25} W H z^{-1} sr^{-1}$ dividing line between FR Is and IIs identified by Fanaroff and Riley (1974, FR). Further we considered the placement of these objects on the radio/optical luminosity diagram (Ledlow and Owen 1996). While FR Is fall as far as 3 orders of magnitude in radio power below the FR I/II break, all three of these transition sources fall about 1 order of magnitude in radio luminosity below the FR I/II break in the radio/optical luminosity diagram of Ledlow and Owen. We also looked at the ratio r of the distance between hot spots to the overall source extent (the original FR I/II defining property). Both 1231+674 and 1433+553 are close to the dividing line of 0.5, although 3C449 has a clear FR I r value. (Table 1). In looking at the FR I/II differences, it is also important to note that FR Is are also not a homogeneous class, and a more careful characterization of their structures is probably in order. For example, although we and others refer to the head-tail source NGC1265 as a prototypical FR I, the Parma et al., 1987 sample of low luminosity sources showed that about one third of them have jets embedded in lobes which is more reminiscent of FR II structures.

A variety of explanations have been proposed for the differences between FR Is and FR IIs, (e.g., Bridle 1986, Williams 1991, De Young 1993, Baum et al., (1995)), *e.g.*, the relative density of the jet, the importance of turbulence, the jet speed and momentum flux, environmental effects of the ISM or winds, effects tied to the optical luminosity, and differences in the central engines. Intermediate values of such properties, or certain environmental conditions, might lead to transition-like objects, but this needs further study. In WATs, the role of the external environment is likely to be important, as shown *e.g.*, by Burns et al., 1994a and Burns et al., 1994b. A slightly different approach to this is taken by O'Donoghue et al., 1996 who were motivated by their previous study of WATs (O'Donoghue et al., 1993) to select non-bent sources in WAT-type environments with intermediate properties similar to WATs in order to study the FR I/II division without the added complication of the bending.

Another interesting possibility is an evolutionary connection between FR Is and FR IIs. For example, Baum et al., (1995) speculate that FR IIs may evolve into FR Is as the accretion rate of the black hole slows down. Owen and Ledlow (1994) suggest that the relation between the FR I/II luminosity transition and the optical luminosity can be explained by an evolutionary connection between the two classifications. Bicknell and Quinn (1994) suggest that FR II activity may be initiated by a merger and that the activity lasts for approximately the dynamical lifetime of the merger and is followed by FR I activity. Furthermore, in this scenario transition objects are sources that were FR IIs and are currently evolving into FR Is.

## 5. Conclusions and Summary

Our two frequency analysis of the WATs 1231+674 and 1433+553 reveal that each source is comprised of two components, a flat spectrum jet and a steep spectrum sheath. There is little spectral variation of either component up to tens of kpc from the nucleus. The flat jet also tends to be less polarized than the steep sheath.

In a future paper, we will present polarization tomography; similar to the spectra tomography, it helps to deconvolve overlapping features. Our preliminary results show that polarization tomography confirms the existence of two separate components in these WATs.

The physical connection between the jets and sheaths is unclear. FR Is were previously thought not to inflate sheaths around their jets; the steep spectrum tails were understood simply as expanded downstream jets. The sheath may be made up of older material that has expanded or diffused from the jet, or a completely separate population, perhaps originating in the ICM; none of the simple models are completely statisfactory. The existence of the sheath must be incorporated into structural and dynamical models of WATs. The WATs are transition objects between FR Is and FR IIs; therefore, the presence of the sheath in these sources may help us understand the nature of the flow patterns and structures in all of these sources.

Our analyses depended on the interpretation of pre-existing data. We would like to thank Jean Eilek and Frazer Owen for contributing their data on 1231+674 and 1433+553. This research was supported, in part, by the Naval Research Laboratory, Naval Academy Research Council and the ONR grant N0001497WR20008 and in part, at the University of Minnesota through NSF grant AST-93-18959 and AST-96-16984.

Table 1.   Source Characteristics

| Source | Redshift | $L_{178}$ (x $10^{25}$ W Hz$^{-1}$ sr$^{-1}$) | r |
|--------|----------|------------------------------------------------|---|
| 1231+674 | 0.1602 | 0.7[1] | 0.4 |
| 1433+553 | 0.1396 | 0.4 | 0.5 |
| 3C449 | 0.0181 | 0.3 | <0.2 |

[1]From $\lambda$ 20 cm data assuming $\alpha$=-0.5. Throughout this paper we have defined the spectral index, $\alpha$, such that $I(\nu) = I_0 \nu^\alpha$.

Throughout the paper, we assume $H_0 = 75\ km\ s^{-1} Mpc^{-1}$.

References for the redshifts and luminosities are as follows: For 1231+674 and 1433+553, O'Donoghue et al. (1990). For 3C449, Sandage 1967 and Katz-Stone and Rudnick 1997b.

Table 2.   New Observations of 1231+674 and 1433+553: Integration Times (hours)

| VLA Configuration | 20 cm | 6 cm | 3.6 cm |
|---|---|---|---|
| C | 0 | 0 | 3 |
| D | 0.5 | 1 | 4 |

Table 3.   Flux Calibrator 3C286

| Observing Band | Frequency (MHz) | Assumed Flux (Jy) |
|:---:|:---:|:---:|
| **20 cm** | | |
| | 1464.9 | 14.6353 |
| | 1514.9 | 14.3995 |
| **6 cm** | | |
| | 4885.1 | 7.5299 |
| | 4835.1 | 7.5782 |
| **3.6 cm** | | |
| | 8414.9 | 5.2799 |
| | 8464.9 | 5.2585 |

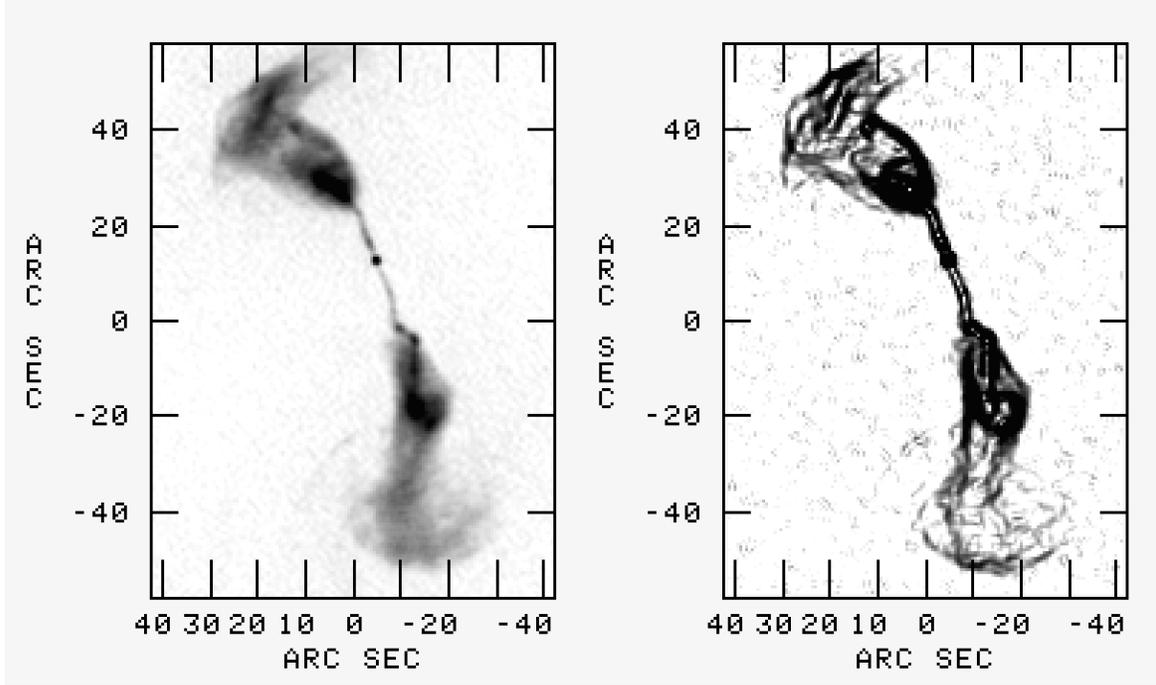

Fig. 1.— High Resolution Images of 1231+674. On the left is the total intensity image at λ 20 cm at a resolution of 1.″4 and peak flux of 5.6 mJy/beam. On the right is an edge-filtered version of this map. The coordinate (0,0) corresponds to RA= 14h33m53.56s, Dec= 55d20'54.9"

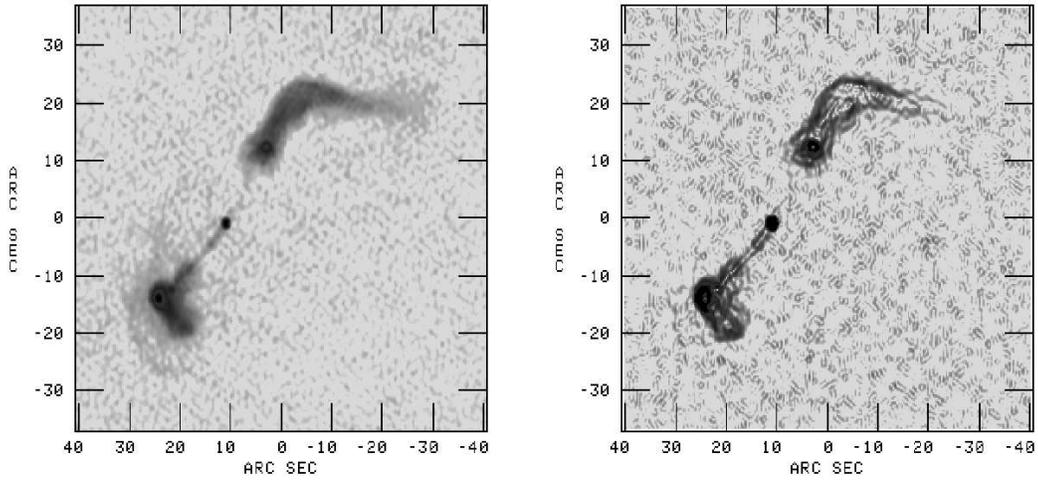

Fig. 2.— High Resolution Images of 1433+553. On the left is the total intensity image at $\lambda$ 20 cm at a resolution of 1.″13 $\times$ 0.″98 at 0° and a peak flux of 10.6 mJy/beam. On the right is an edge-filtered version of this map. The coordinate (0,0) corresponds to RA= 12h31m03.08s, Dec= 67d24'08.2".

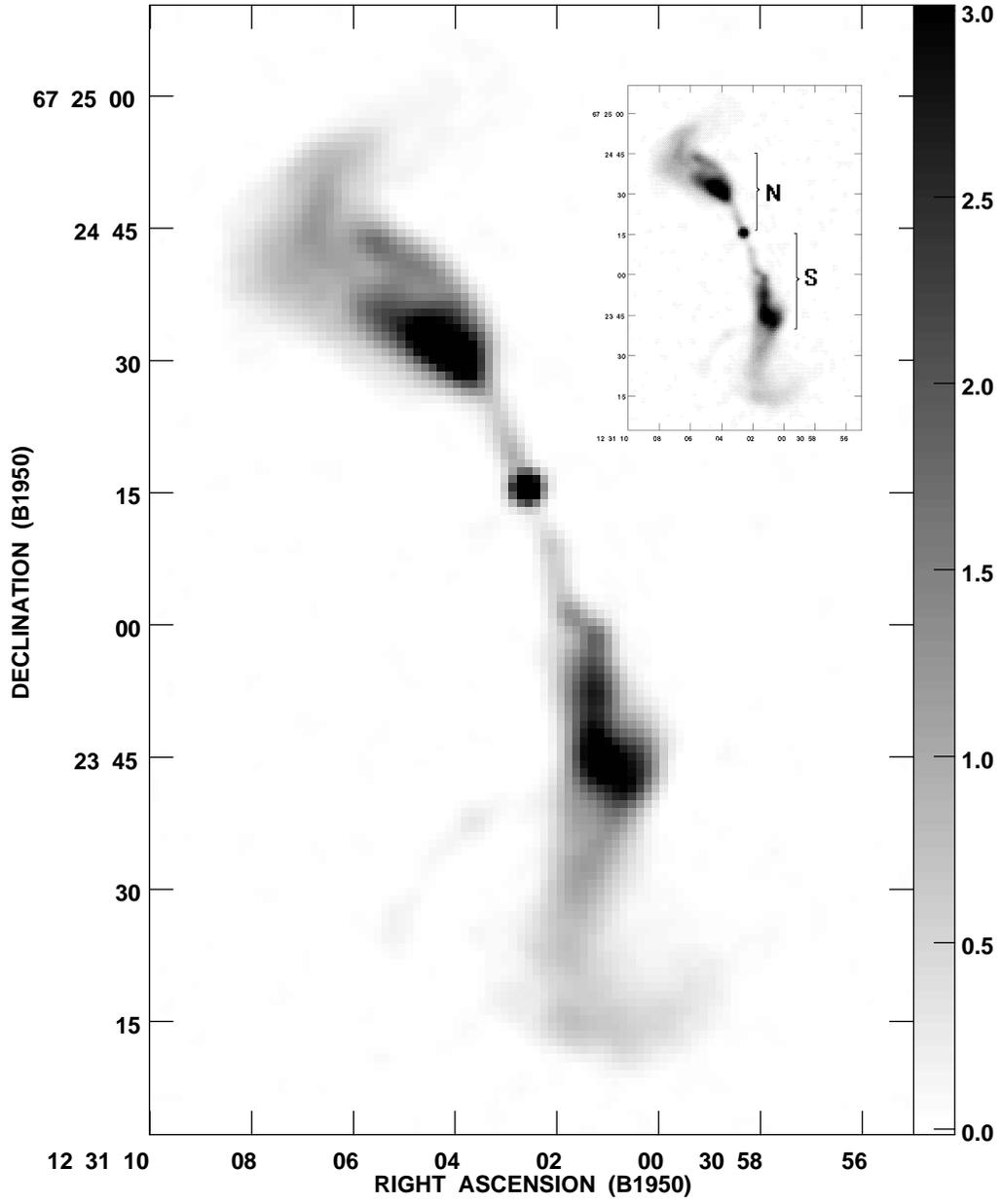

Fig. 3.— Total intensity map of 1231+674 at λ 3.6 cm; resolution = 2".7.

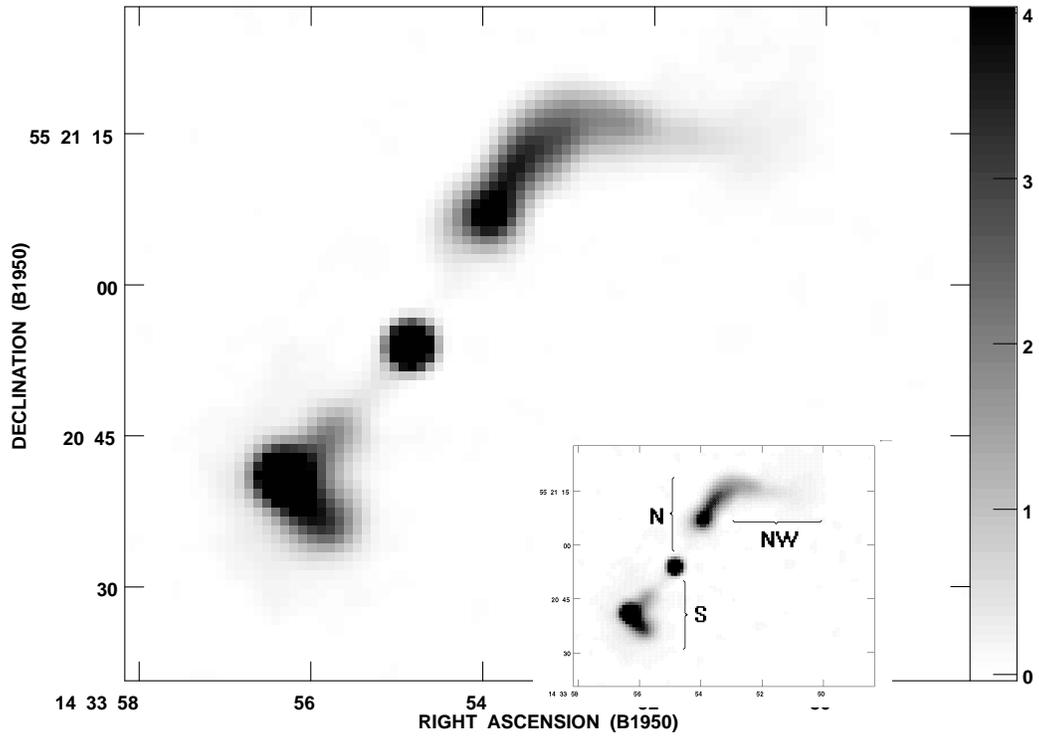

Fig. 4.— Total intensity map of 1433+553 at $\lambda$ 3.6 cm; resolution = 2".9.

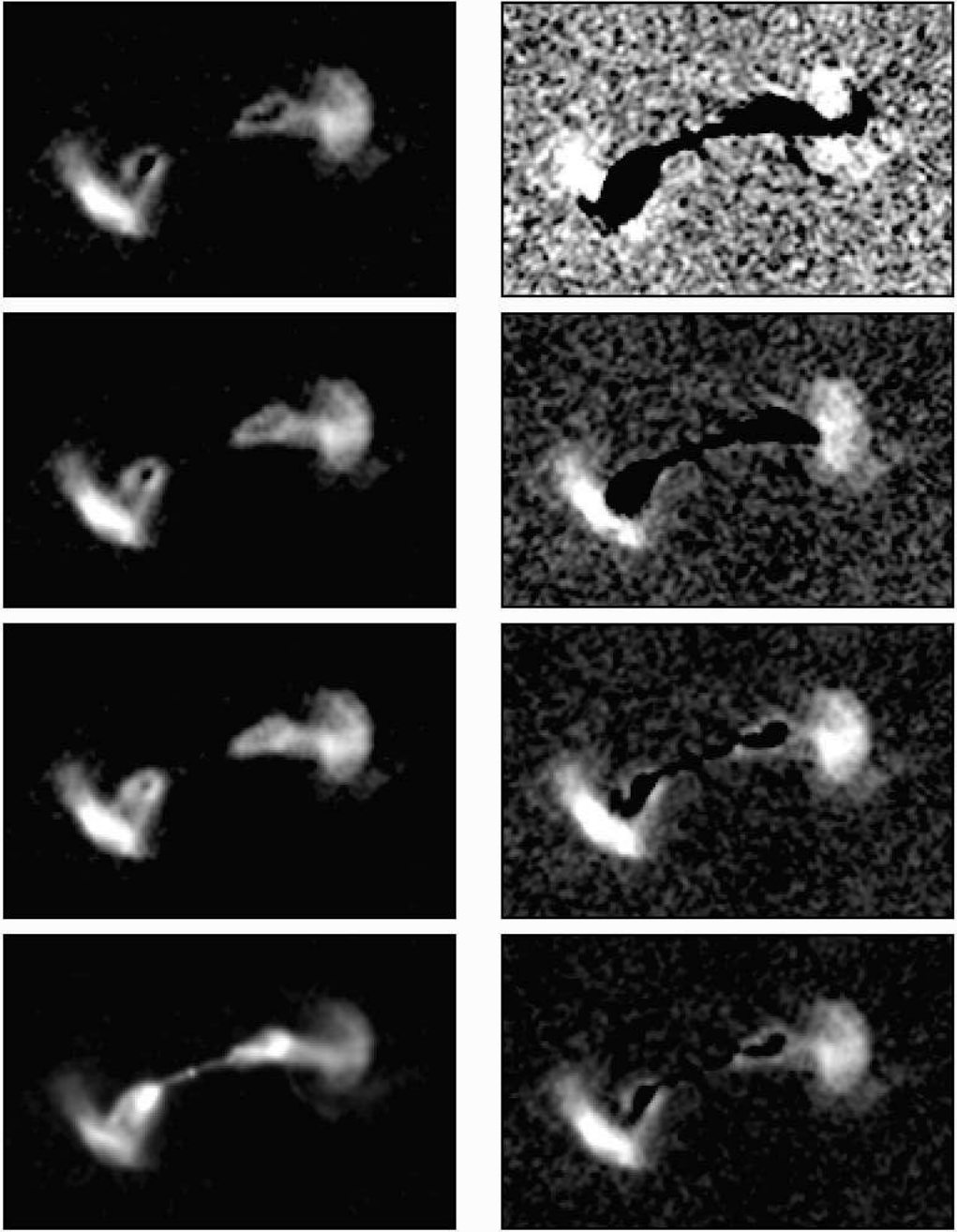

Fig. 5.— Spectral tomography gallery of 1231+674. The top left image is the total intensity at $\lambda$ 20 $cm$. Then from left to right are tomography (difference) maps with $\alpha_t$ equals -0.55, -0.60, -0.65, respectively. On the bottom, $\alpha_t$ equals -0.70, -0.75, -0.85, and -1.20 respectively. (Images are in landscape orientation.)

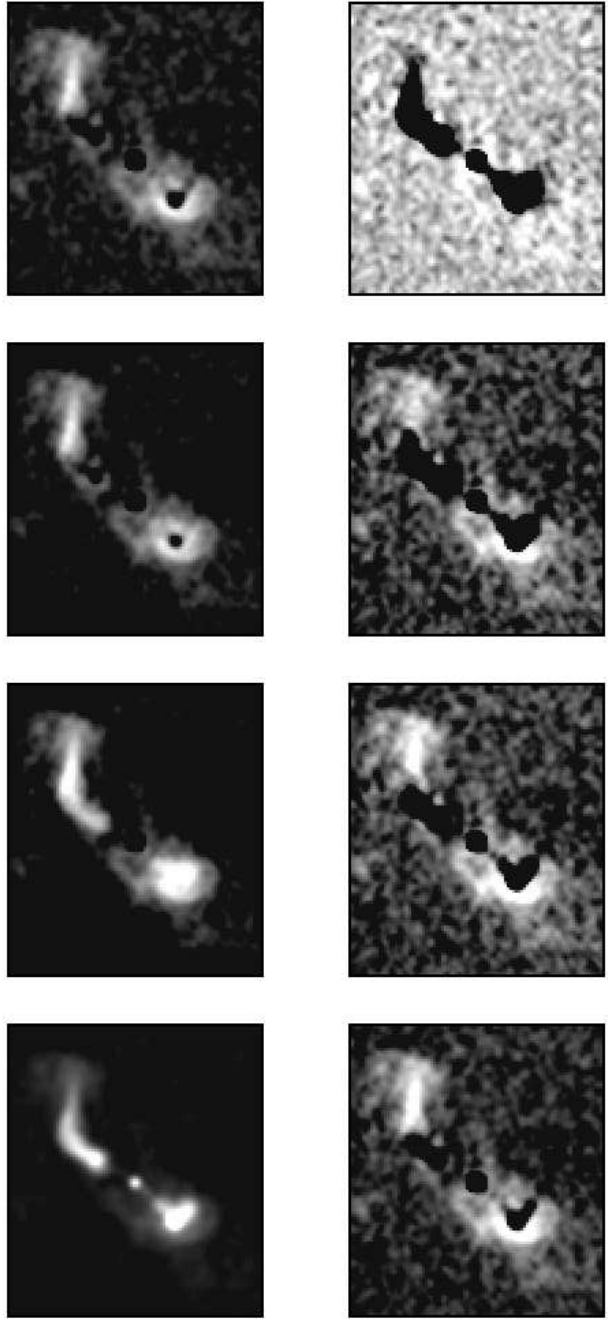

Fig. 6.— Spectral tomography gallery of 1433+553. The top left image is the total intensity at $\lambda$ 20 $cm$. Then from left to right are tomography (difference) maps with $\alpha_t$ equals -0.55, -0.65, -0.70, respectively. On the bottom, $\alpha_t$ equals -0.75, -0.85, -0.95, and -1.15 respectively. (Images are in landscape orientation.)

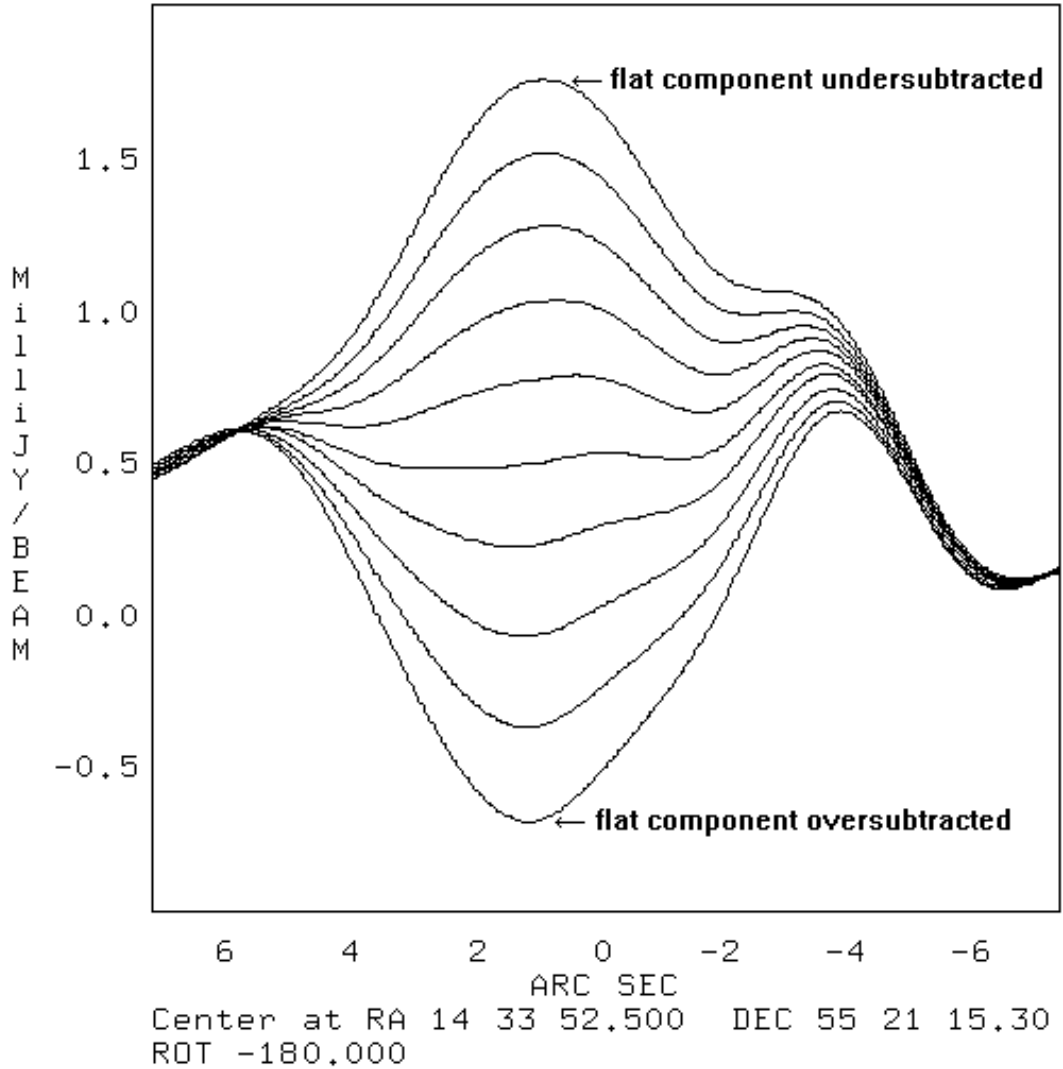

Fig. 7.— Slice through a series of spectral tomography maps of 1433+553. The range in $\alpha_t$ is from -0.72 (top) to -0.9 (bottom); the increment in $\alpha_t$ is 0.02. The flat component is at approximately +2" from the center of these slices. The flat component disappears from the middle slice, implying to a spectral index of $\approx$ -0.82.

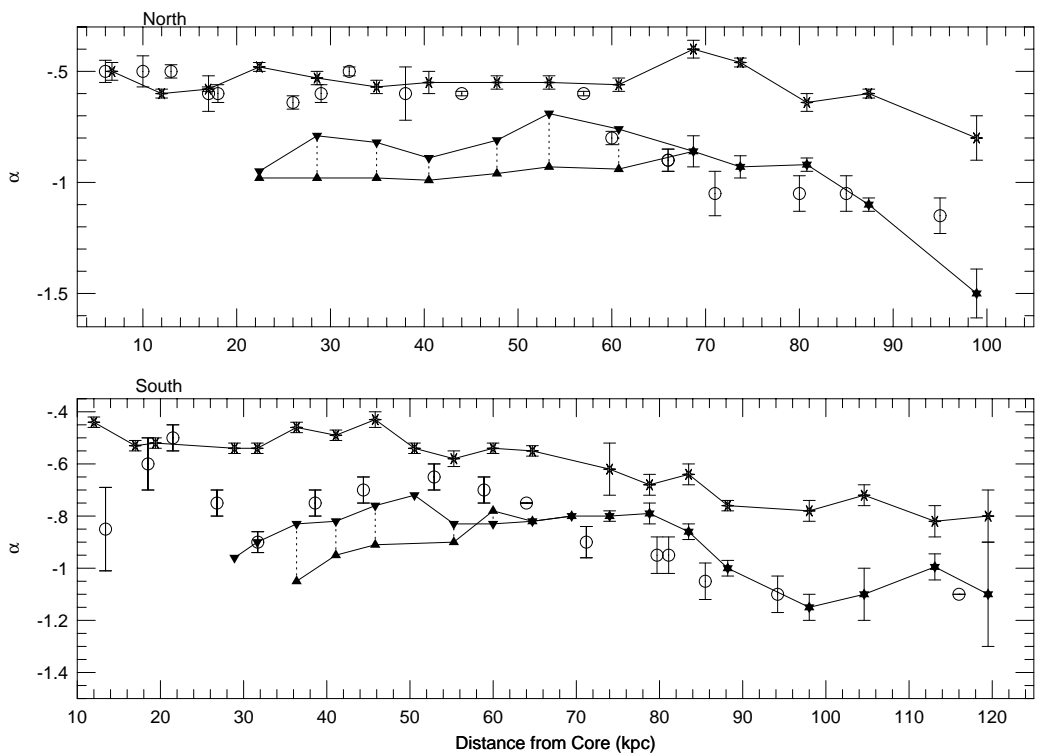

Fig. 8.— Spectral index as a function of distance from the core for 1231+674. The stars (triangles) represent the flat jet (sheath); the downward (upward) pointing triangles are from the east (west) side of the sheath. The spectral indices of the flat jet and the sheath were measured between $\lambda\lambda$ 20 and 3.6 cm. The open circles are taken from O'Donoghue et al. (1990); they are from integrations of the surface brightness over slices taken at $\lambda\lambda$ 20 and 6 cm.

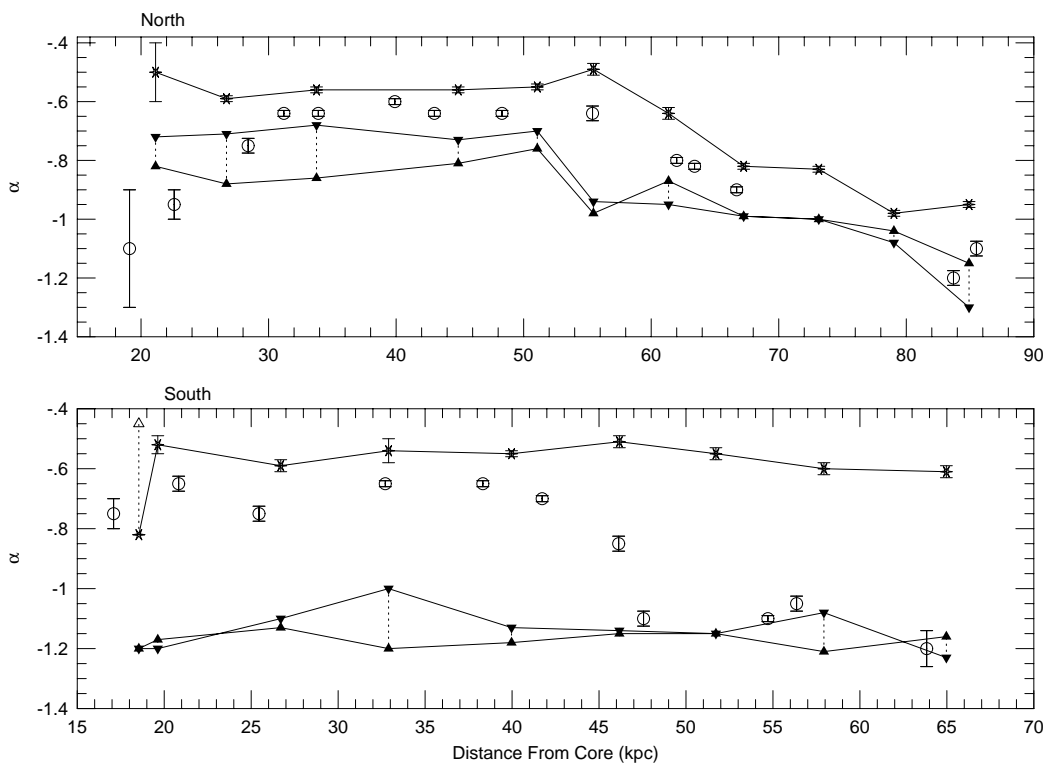

Fig. 9.— Spectral index as a function of distance from the core for 1433+553. The stars (triangles) represent the flat jet (sheath); the downward (upward) pointing triangles are from outside (inside) of the flat jet. The spectral indices of the flat jet and the sheath were measured between $\lambda\lambda$ 20 and 3.6 cm. The open circles are taken from O'Donoghue et al. (1990); they are from integrations of the surface brightness over slices taken at $\lambda\lambda$ 20 and 6 cm.

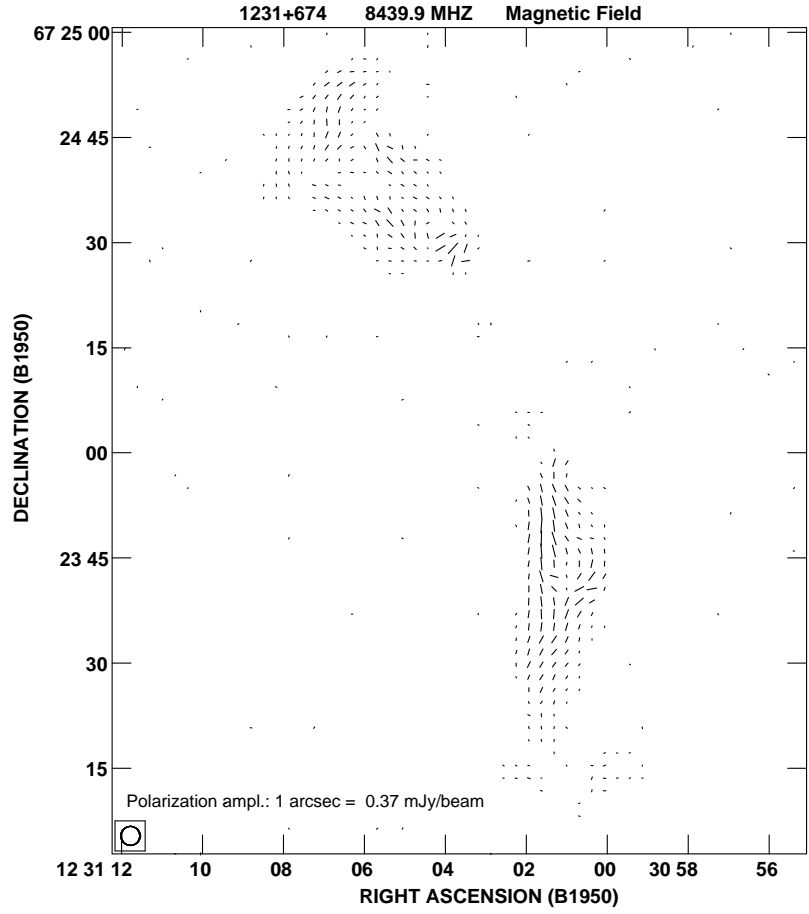

Fig. 10.— Vectors showing the strength of the polarized flux at $\lambda$ 3.6 *cm* and the direction of the inferred magnetic field, as described in the text.

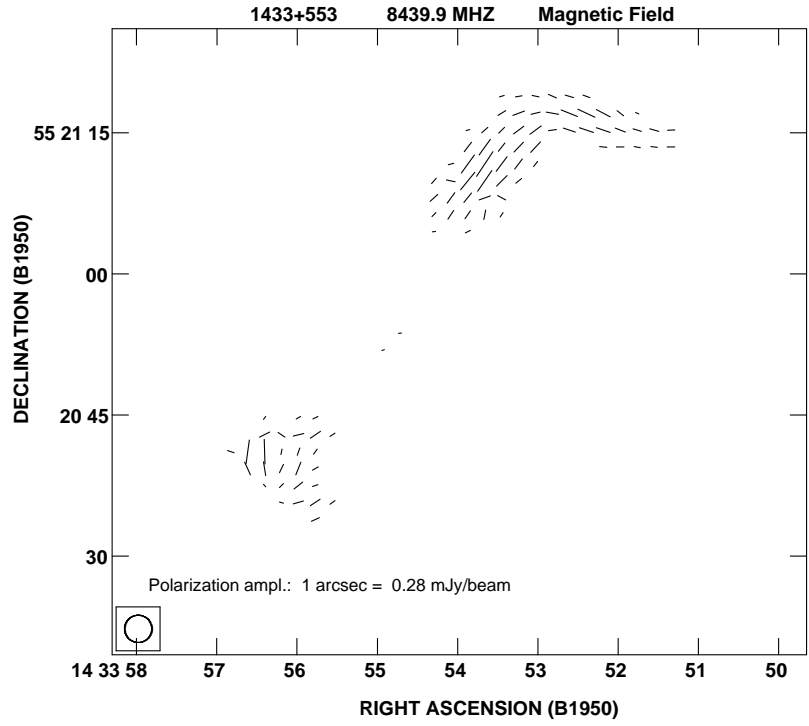

Fig. 11.— Vectors showing the strength of the polarized flux at $\lambda$ 3.6 $cm$ and the direction of the inferred magnetic field, as described in the text.

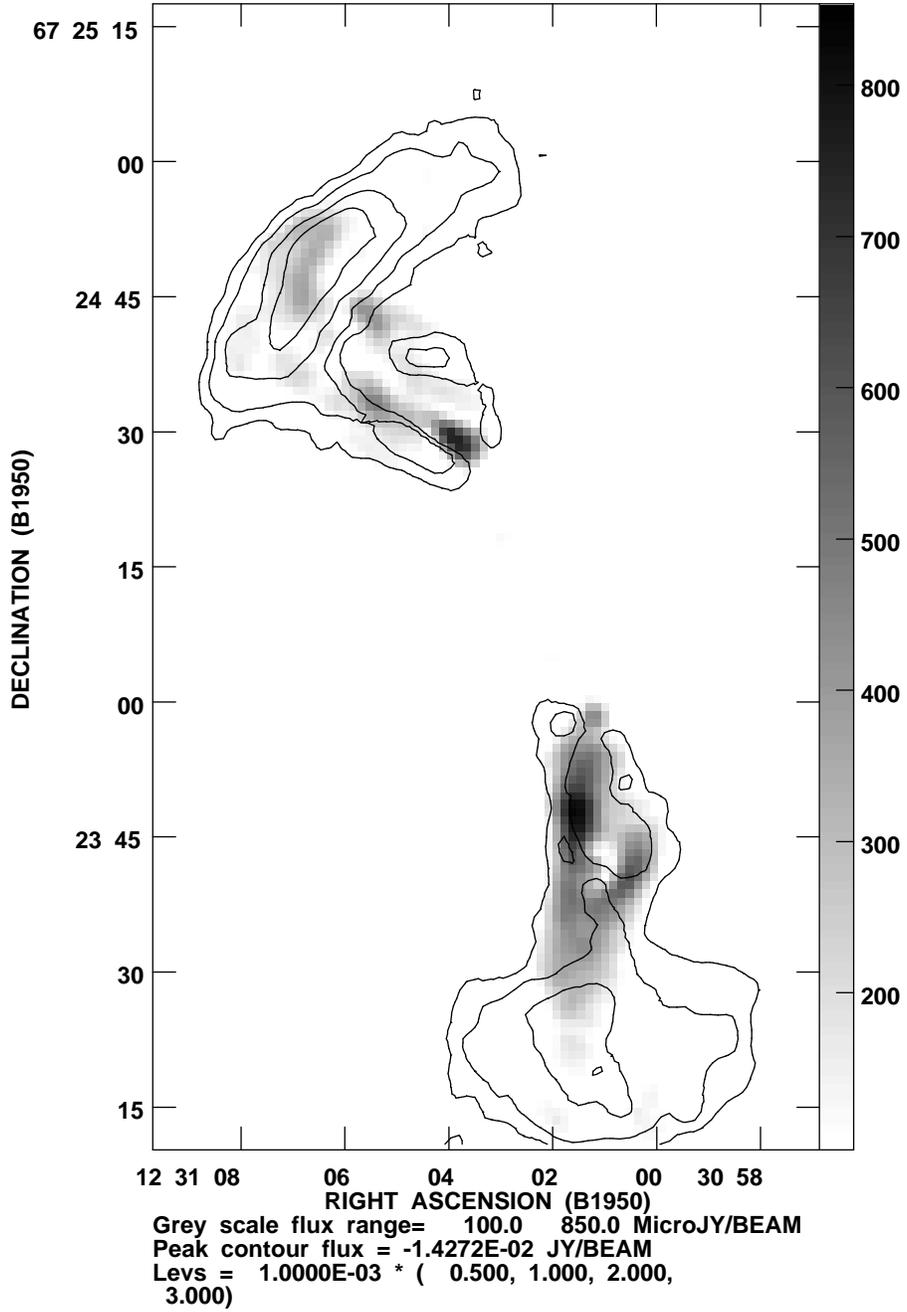

Fig. 12.— An overlay of a spectral tomography map (contours) and the polarization intensity map (greyscale) of 1231+674.

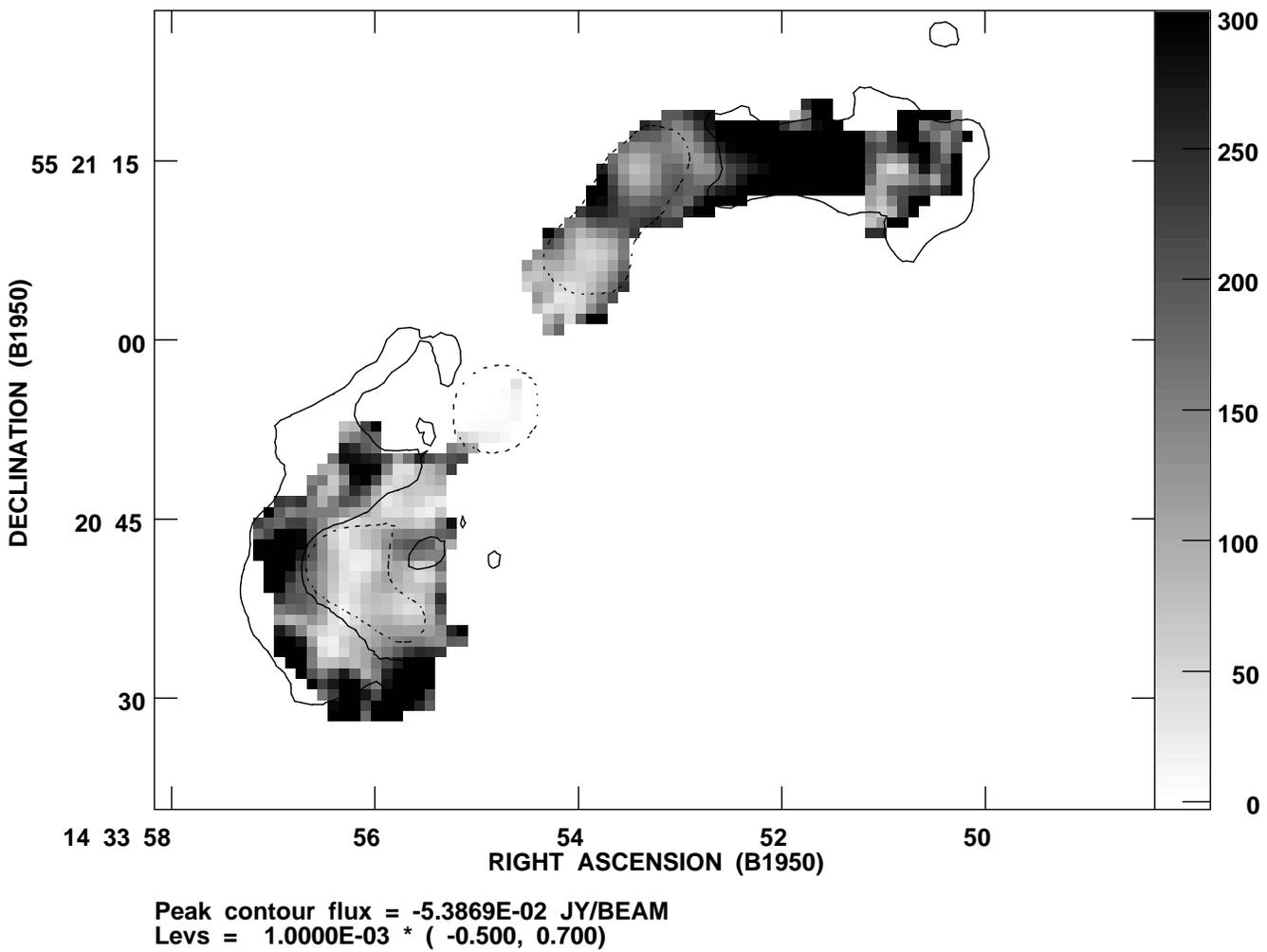

Fig. 13.— An overlay of a spectral tomography map (contours) and the fractional polarization map (greyscale) of 1433+553.

Peak contour flux = -5.3869E-02 JY/BEAM
Levs = 1.0000E-03 * ( -0.500, 0.700)